\documentclass[iop,apj,numberedappendix]{emulateapj}
\usepackage{graphicx}
\usepackage{latexsym}
\usepackage{amsfonts,amsmath,amssymb}
\usepackage{url}
\usepackage[utf8]{inputenc}
\usepackage{hyperref}
\hypersetup{colorlinks=false,pdfborder={0 0 0}}
\usepackage{textcomp}
\usepackage{longtable}
\usepackage{multirow,booktabs}
\usepackage[english]{babel}
\usepackage{natbib}
\usepackage[usenames]{color}

\usepackage{natbib}

\def\sgra{Sgr~A$^{\ast}$}

\def\lsim{\mathrel{\raise.3ex\hbox{$<$\kern-.75em\lower1ex\hbox{$\sim$}}}}
\def\gsim{\mathrel{\raise.3ex\hbox{$>$\kern-.75em\lower1ex\hbox{$\sim$}}}}
\def\gtwid{\mathrel{\raise.3ex\hbox{$>$\kern-.75em\lower1ex\hbox{$\sim$}}}}
\def\proptwid{\mathrel{\raise.3ex\hbox{$\propto$\kern-.75em\lower1ex\hbox{$\sim$}}}}

\defcitealias{Johnson_Gwinn_2015}{JG15}

\shorttitle{The Intrinsic Shape of Sagittarius A* at 3.5-mm Wavelength}
\shortauthors{Ortiz-Le{\'{o}}n et al.}

\begin{document}


\title{The Intrinsic Shape of Sagittarius A* at 3.5-mm Wavelength}


\author{Gisela N.~Ortiz-Le{\'{o}}n\altaffilmark{1},
Michael D.~Johnson\altaffilmark{2},
Sheperd S.~Doeleman\altaffilmark{2,3},
Lindy Blackburn\altaffilmark{2},
Vincent L.~Fish\altaffilmark{3},
Laurent Loinard\altaffilmark{1,4},
Mark J.~Reid\altaffilmark{2},
Edgar Castillo\altaffilmark{5,6},
Andrew A.~Chael\altaffilmark{2},
Antonio Hern{\'{a}}ndez-G{\'{o}}mez\altaffilmark{1},
David Hughes\altaffilmark{5},
Jonathan Le{\'{o}}n-Tavares\altaffilmark{5,7},
Ru-Sen Lu\altaffilmark{4},
Alfredo Monta\~na\altaffilmark{5,6},
Gopal Narayanan\altaffilmark{8},
Katherine Rosenfeld\altaffilmark{2},
David S{\'{a}}nchez\altaffilmark{5},
F.~Peter Schloerb\altaffilmark{8},
Zhi-qiang Shen\altaffilmark{9},
Hotaka Shiokawa\altaffilmark{2},
Jason SooHoo\altaffilmark{3}, and
Laura Vertatschitsch\altaffilmark{2}
}

\altaffiltext{1}{Instituto de Radioastronom{\'{i}}a y Astrof{\'{i}}sica, Universidad
Nacional Aut{\'{o}}noma de M{\'{e}}xico, Morelia 58089, M{\'{e}}xico}
\altaffiltext{2}{Harvard-Smithsonian Center for Astrophysics, 60 Garden
Street, Cambridge, MA 02138, USA}
\altaffiltext{3}{Massachusetts Institute of Technology, Haystack Observatory,
Route 40, Westford, MA 01886, USA}
\altaffiltext{4}{Max Planck Institut f\"ur Radioastronomie, Auf dem H\"ugel
69, D-53121 Bonn, Germany}
\altaffiltext{5}{Instituto Nacional de Astrof{\'{i}}sica {\'{O}}ptica y
Electr{\'{o}}nica, Apartado Postal 51 y 216, 72000 Puebla, M{\'{e}}xico}
\altaffiltext{6}{CONACyT Research Fellow}
\altaffiltext{7}{Sterrenkundig Observatorium, Universiteit Gent, Krijgslaan 281-S9, B-9000 Gent, Belgium}
\altaffiltext{8}{Department of Astronomy, University of Massachusetts,
Amherst, MA 01002, USA}
\altaffiltext{9}{Shanghai Astronomical Observatory, 80 Nandan Road, Shanghai 200030, China }

\email{g.ortiz@crya.unam.mx}

\begin{abstract}
The radio emission from \sgra\ is thought to be powered by accretion onto a supermassive black hole of $\sim\! 4\times10^6~ \rm{M}_\odot$ at the Galactic Center.  At millimeter wavelengths, Very Long Baseline Interferometry (VLBI) observations can directly resolve the bright innermost accretion region of \sgra. Motivated by the addition of many  sensitive, long baselines in the north-south direction, we developed a full VLBI capability  at the Large Millimeter Telescope Alfonso Serrano (LMT).  We successfully detected \sgra\ at 3.5~mm with an array consisting of 6 Very Long Baseline Array telescopes and the LMT.
We model the source as an elliptical Gaussian brightness distribution and estimate the  scattered size and orientation of the source from closure amplitude and self-calibration analysis, obtaining consistent results between methods and epochs. 
We then use the known scattering kernel to  determine the intrinsic two dimensional source size at 3.5 mm: $(147\pm7~\mu\rm{as}) \times (120\pm12~\mu\rm{as})$, at position angle $88^\circ\pm7^\circ$ east of north.  
Finally, we detect non-zero closure phases on some baseline triangles, but we show that these are consistent with being introduced by refractive scattering in the interstellar medium and do not require intrinsic source asymmetry to explain.
\end{abstract}


\section{Introduction}

The compact radio source Sagittarius~A$^\ast$ (\sgra) at the center of the Galaxy is associated with a supermassive black hole of ${\sim}\, 4\times10^6~ \rm{M}_\odot$ \citep{Ghez_2008, Gillessen_2009}. The mechanism responsible for the radio emission is thought to be synchrotron from a jet-like outflow \citep{Markoff_2007,Falcke_2009}, a radiatively inefficient accretion flow onto the black hole \citep[e.g.,][]{Narayan_1995,Yuan_2003,Broderick_2009} or an almost isothermal jet coupled to an accretion flow \citep{Moscibrodzka_2013}. Different jet and accretion disk models can be tested by modeling the radio through submillimeter spectrum of \sgra\ \citep[e.g.,][]{Markoff_2007},  the frequency-dependent source size \citep[e.g.,][]{Bower_2004,Moscibrodzka_2013,Chan_2015}, and  data from millimeter Very Long Baseline Interferometry (VLBI) observations \citep{Dexter_2012,Broderick_2011}. 

At wavelengths longer than a few centimeters, the image of \sgra\ is heavily scattered by the intervening ionized interstellar medium, and the scattering determines the size of the measured image. The effect of this scattering decreases at shorter wavelengths, with a $\lambda^2$ dependence, and 
 VLBI observations at wavelengths shorter than a centimeter have found deviations from the $\lambda^2$ law, suggesting that intrinsic source structure contributes to the apparent image at these wavelengths \citep{Doeleman_2001,Bower_2004,Shen_2005,Bower_2006,Krichbaum_2006}. The intrinsic two-dimensional source size can then be estimated by extrapolating the scattering properties from longer wavelengths and then deconvolving the scattering ellipse from the observed size. At a wavelength of one millimeter or less, the scatter-broadening is subdominant to intrinsic structure in the image \citep{Doeleman_2008,Fish_2011,Johnson_2015}. 

Because of the lack of good north-south baselines in existing VLBI arrays, efforts to study the intrinsic structure of \sgra\ at 3.5~mm have been mostly limited to the east-west direction. To unambiguously determine the intrinsic two-dimensional structure of \sgra, VLBI observations with higher angular resolution in the north-south direction are needed. In this paper, we describe such observations of \sgra\ obtained at $\lambda$ = 3.5~mm with the National Radio Astronomy Observatory\footnote[1]{The National Radio Astronomy Observatory (NRAO) is a facility of the  National Science Foundation operated under cooperative agreement by Associated Universities, Inc.} Very Long Baseline Array (VLBA) and the Large Millimeter Telescope Alfonso Serrano (LMT) located in Central Mexico, operated in concert as a single VLBI array. This required that the LMT be equipped as a VLBI station, as we now describe.

\section{VLBI at the LMT}

\begin{figure*}[t]
\begin{center}
\includegraphics[width=0.8\textwidth]{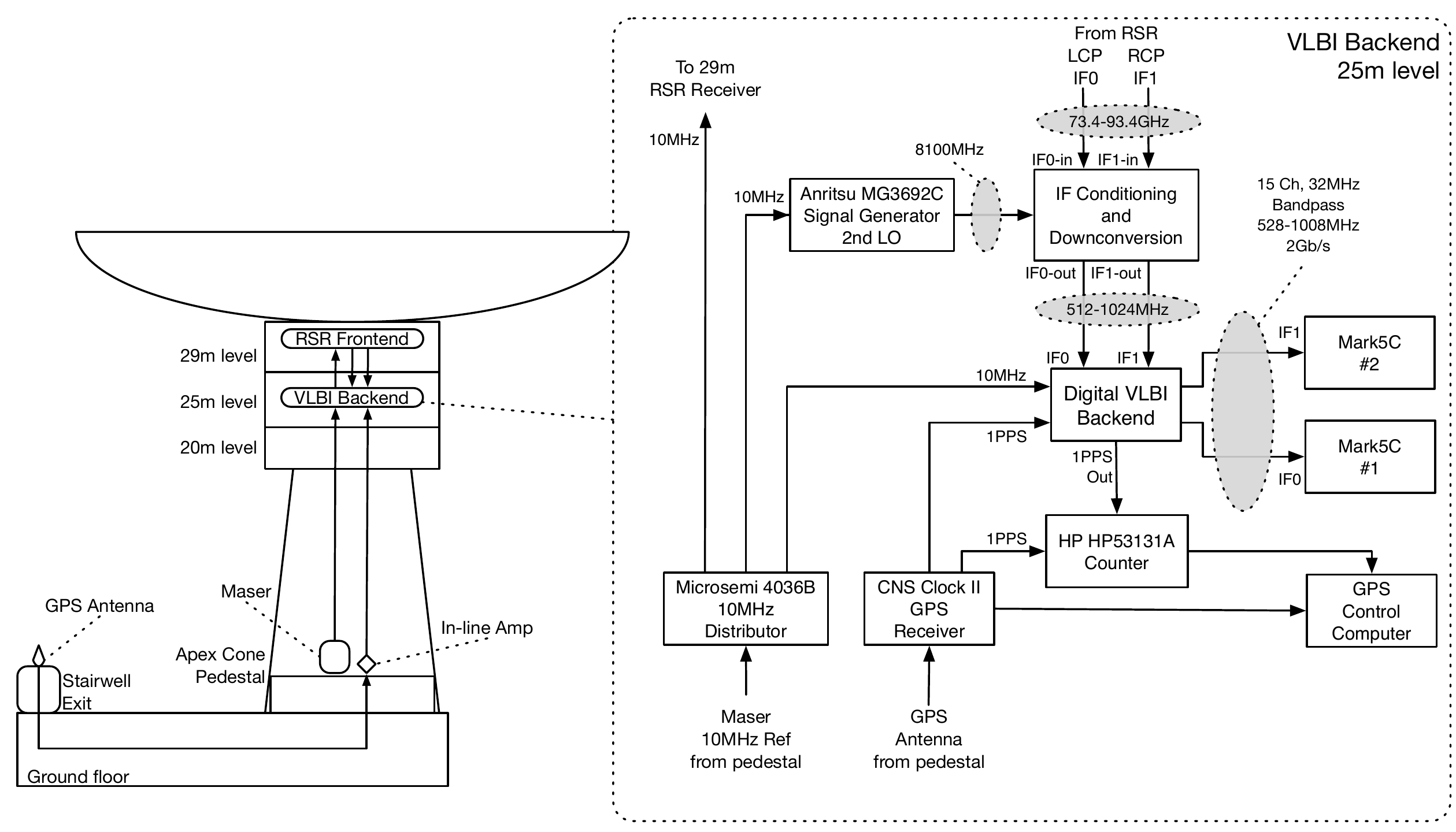}
\caption{\label{fig:blkdiag} Block diagram of VLBI instrumentation setup at the LMT for the April 2015 observations. 
}
\end{center}
\end{figure*}

Situated at an altitude of 4,600 meters at the summit of Volc\'an Sierra Negra in Central Mexico, 
the  LMT has a large collecting area (\mbox{32-m} circular aperture currently operational, extending to the full 50-m diameter by 2017) and geographical location that make it particularly useful for mm-wavelength VLBI observations.  Technical work leading to development of VLBI capability at the LMT was the product of a multi-year collaboration between Instituto Nacional de Astrof{\'{i}}sica, {\'{O}}ptica y Electr{\'{o}}nica (INAOE), University of Massachussets (UMass), Smithsonian Astrophysical Observatory (SAO), Massachussets Institute of Technology (MIT) Haystack Observatory, Universidad Nacional Aut{\'{o}}noma de M{\'{e}}xico (UNAM), and NRAO.   Recognizing the importance of LMT participation in 3.5~mm VLBI networks (e.g., the VLBA or the Global Millimeter VLBI Array -- GMVA), and in the Event Horizon Telescope (EHT) project at 1.3-mm wavelength, these groups began planning VLBI tests in 2012.  First 3.5-mm observations were scheduled in April 2013, for which a full VLBI recording system was installed.  This included integration at the Sierra Negra site of:
\begin {itemize}
\item A GPS receiver (model CNS) to enable synchronization with other VLBI sites.
\item A custom built Radio Frequency (RF) downconverter to shift the output of the facility Redshift Search Receiver (RSR) to a standard VLBI intermediate frequency (IF) range of 512-1024 MHz.
\item A digital backend to digitize and format data for VLBI recording \citep{Whitney_2013}.
\item Two high-speed hard-disk Mark5c VLBI recorders.\footnote[2]{http://www.haystack.edu/tech/vlbi/mark5/mark5$\_$memos/057.pdf \\ Unlike nominal operations of the VLBA, the LMT did not record dual polarization on the same disc set.}
\end{itemize} 

The RSR is one of the two instruments currently available at the LMT. The RSR has two H and V linear polarization receivers  that instantaneously cover a wide frequency range of 73 -- 111 GHz, and has a dedicated backend spectrometer that covers the entire band with a spectral resolution of 31 MHz \citep{erickson_2007}. The receivers  are chopped between the ON and OFF source positions (beam 1 and 0, respectively) separated by $76''$.  Sources are tracked on beam 1 during VLBI observations. 
The RSR has 2 fixed first local oscillators (LOs) at 93.4 and 112.3 GHz, which are used to downconvert the frontend band into two 0 -- 20 GHz IF bands. For this VLBI experiment, we used the 73 -- 93.4 GHz band (see Figure \ref{fig:blkdiag}) for further down-conversion and processing.

A hydrogen maser, typically used to provide a stable frequency reference for VLBI, was not available for the 2013 observations, so an ultra-stable quartz crystal oscillator loaned by the Applied Physics Laboratories of Johns Hopkins University was used.  This unit has an Allan Deviation of $<\! 10^{-13}$ over integration times from 1-10 seconds, resulting in coherence losses of $<\! 25\%$ at 3.5~mm wavelength.  This crystal was thus sufficient for initial tests, but not for scientific observations.  To convert the Linear Polarization natively received by the RSR to Circular Polarization, a quarter-wave plate made of grooved dielectric was inserted into the telescope optics, and for subsequent observations Left Circular Polarization was selected.

Using this test setup (see Figure~\ref{fig:blkdiag}), several SiO maser sources (v=1, J=2-1) and bright AGN were detected on baselines from the LMT to the VLBA, confirming the stability of the LMT RSR and  VLBI system performance.  In 2014, this same setup was augmented by installation and integration of a hydrogen maser frequency standard (manufactured by Microsemi) that is housed in the pedestal room of the telescope.  A custom-built enclosure provides a temperature stable environment for the maser, and a low-noise distribution system installed near the VLBI equipment routes the  maser reference to phase lock all VLBI instrumentation.

\begin{figure}[t]
\begin{center}
\includegraphics[width=0.42\textwidth]{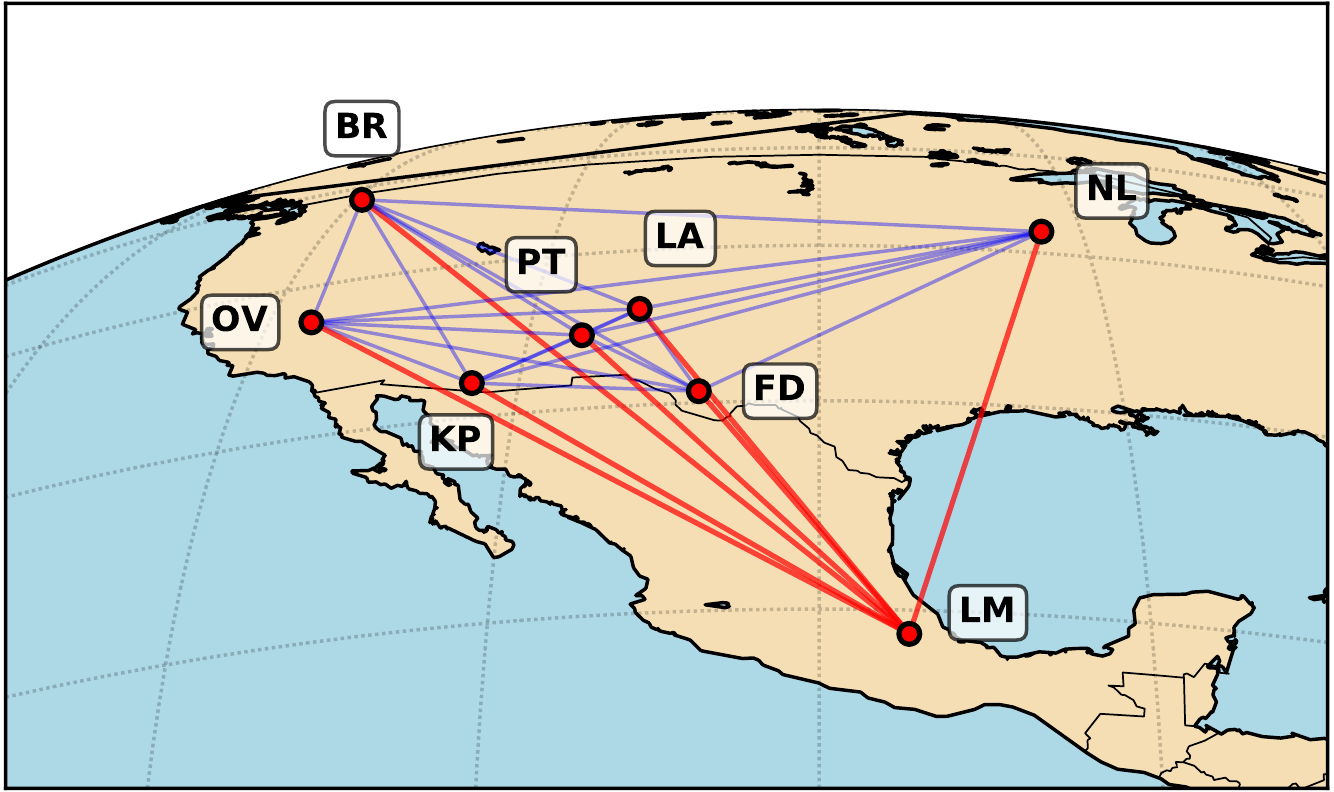}\\ \vspace{0.2cm}
\includegraphics[width=0.47\textwidth]{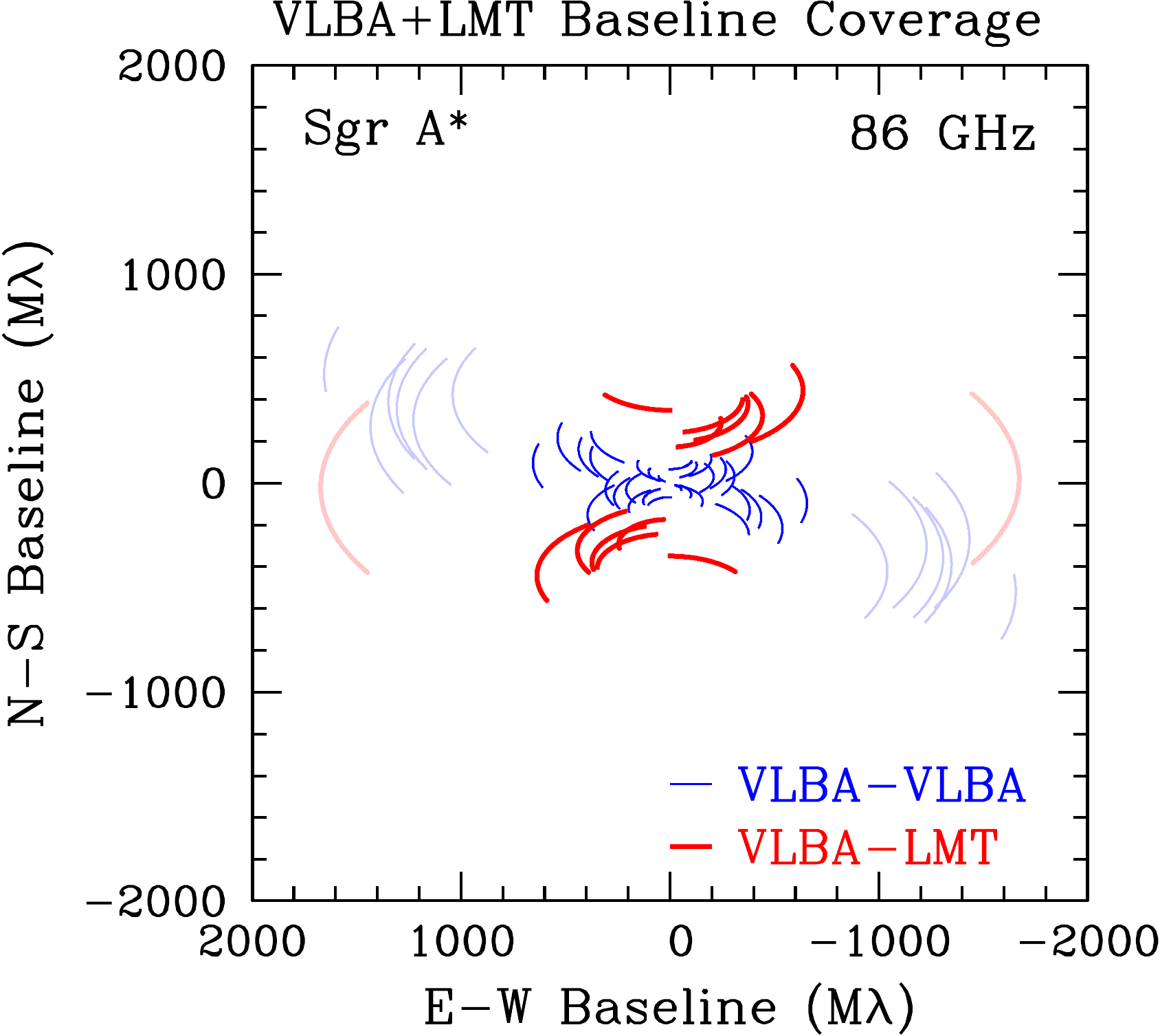}
\caption{\label{fig:LMT-Globe} ({\it Top}) The 3.5 mm stations of the VLBA and the LMT. ({\it Bottom}) Corresponding u-v coverage; the faint tracks denote baselines to Mauna Kea, on which we do not detect \sgra.
}
\end{center}
\end{figure}

Commissioning observations in 2014 were conducted over the course of four nights between the VLBA and the LMT.  A precise position for the LMT was measured by modeling the delays and rates of VLBI detections on strong quasars over a wide range of elevation.  The operational location of the LMT in the International Terrestrial Reference Frame geocentric coordinates is: (X,Y,Z) = $(-7.687156(2)\times10^{5}$ m, $-5.9885071(2)\times10^{6}$ m, $2.0633549(5)\times10^{6}$ m).  This location corresponds to the projection point of the horizontal axis onto the vertical axis.
Figure~\ref{fig:LMT-Globe} shows the VLBA and LMT as seen by \sgra\ and the corresponding baseline coverage. 

\section{Observations and Data Calibration}
\label{sec::Observations}

The observations reported here (project code BD183) were obtained in 2015 by operating the eight VLBA antennas equipped with 3.5~mm receivers and the LMT as a single VLBI array. The central frequency was 86.068 GHz. A total of 27 hours of telescope time were allocated to the project, which were covered in 3 epochs of 9 hours each on 2015 April 24, 27 and 28 (codes A, C and D, respectively). Because scans for pointing and calibration were also included in each observation, only about 3.6 hours were actually spent on-source in each epoch.  Observations were triggered at all sites based on expected weather conditions at LMT and North Liberty, the key stations of the project. Data were recorded at a rate of 2 Gb~s$^{-1}$, and taken in left circular polarization, with 480 MHz of bandwidth covered by 15 32-MHz IF channels. 

In the first epoch, the LMT RSR tracked on the wrong beam (beam 0), and this was caught  just before finalizing observations. 
On the second epoch, the station at Pie Town (PT) experienced precipitation during most of the observing run so data were highly affected. On the third epoch, the Los Alamos (LA) recording system corrupted the data due to timing issues.  Thus, the data taken at the LMT on first epoch, at PT on the second epoch, and at LA on the third epoch were discarded. Because baselines between Mauna Kea (MK) and the rest of the array resolve out the emission from \sgra, the source was not detected on these baselines.
Fringe detections on \sgra\ were therefore obtained with an array consisting of 7 stations in each of the three epochs. 

 
For the remainder of the paper, we will focus on the last two epochs (BD183C and BD183D) because our goal of constraining the intrinsic size of \sgra\ at 3~mm relies heavily on the north-south baselines provided by the LMT.
 
The initial data reduction was done using the Astronomical Image Processing System \citep[AIPS; ][]{Greisen_2003}. Phase calibration was performed as follows. Corrections for the antenna axis offset at the LMT and for voltage offsets in the samplers at all stations  were first applied to the data. Single-band delays were determined by fringe-fitting on a strong calibrator (3C279 for BD183C and  3C454.3 for BD183D), and the solutions were applied to all scans in the corresponding observing night. \sgra\ was then fringe-fitted, producing rate and delay solutions every 1 minute. These solutions were smoothed using a median window filter smoothing function with a 6-minute filter time, and then applied to the data. A single bandpass solution was derived from the auto-correlations on the continuum sources and applied to the data after fringe-fitting. At this point, all scans with non-detections were flagged. Also, the outer 4.5 MHz from the edge of each IF were discarded because these are adversely affected by the bandpass response function.

\begin{figure}[t]
\begin{center}
\includegraphics[width=1\columnwidth]{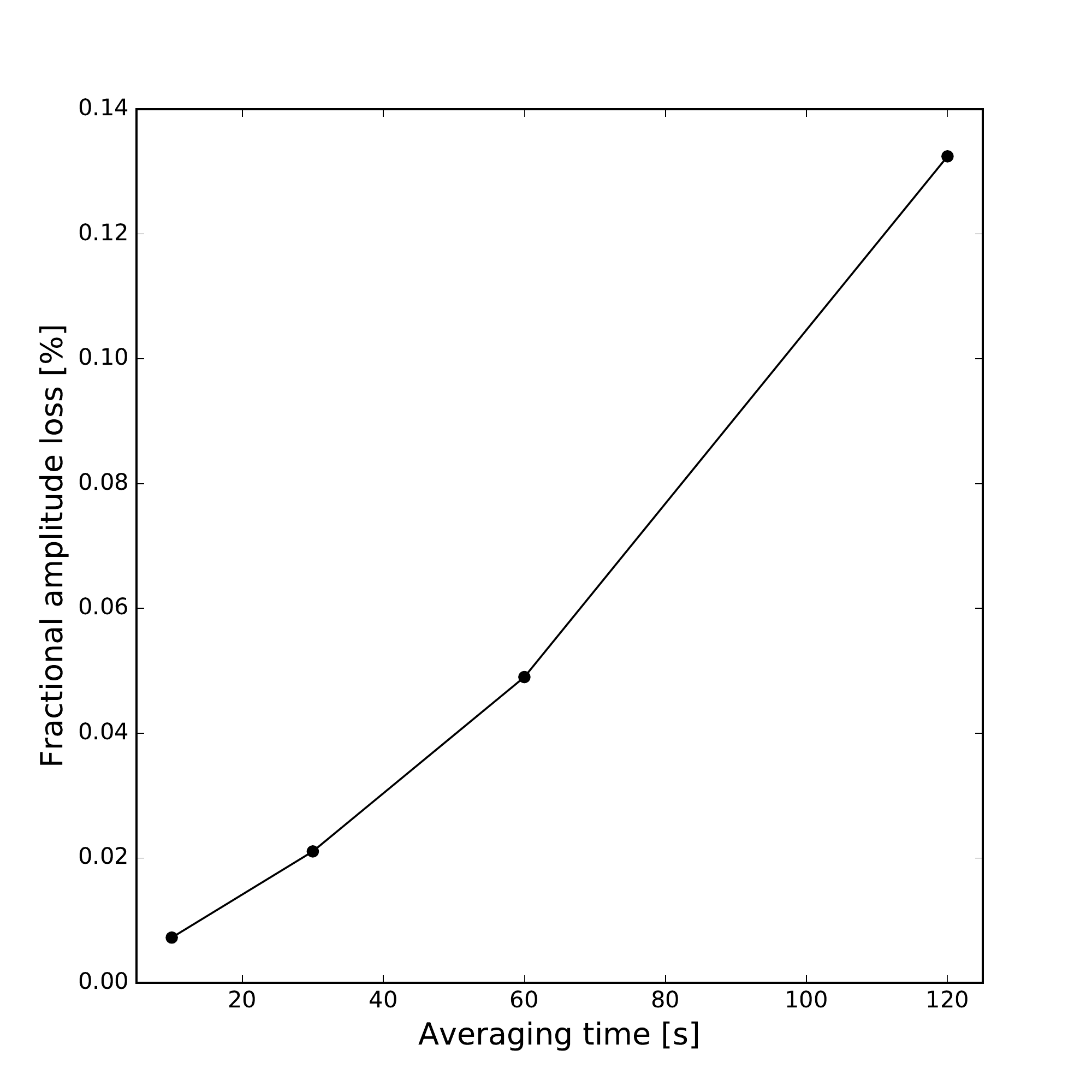}
\caption{\label{fig:loss} Fractional amplitude loss as a function of averaging time for a scan on 3C279 taken in the first epoch. We estimated this fraction for each baseline and took the maximum values to show in the plot. }
\end{center}
\end{figure}

To optimize the coherent averaging of visibilities, we estimated the atmospheric coherence time of our data by examining the ratios of debiased coherent to incoherent averages \footnote[3]{A coherent average takes the vector-average of complex visibilities, preserving the coherence of phase over time and frequency \citep{TMS_2007}.  An incoherent average takes the scalar-average of complex visibilities segmented at short-length times. A debiased average corrects visibility amplitudes by the noise bias introduced because of the inherently positive nature of amplitudes.}  as a function of time using a scan on 3C279.  For every baseline, we found that the fractional amplitude loss is  $<0.7\%$ for $t_{\rm avg} = 10$ seconds (see Figure \ref{fig:loss}).
Considering the fractional amplitud loss scales with the line-of-sight optical depth,  and because  \sgra\ is at lower elevation, we estimated the loss increases to $<4\%$ in the worst case.
To ensure that closure relationships (discussed below) were not affected by coherence losses, we then utilized 10-second coherent averages.  For this segment of time the losses can be considered negligible in all of our data. After this coherent averaging in time and across the full bandwidth, these phase-only calibrated data  were exported as FITS files for further analysis outside of AIPS.  

\section{Analysis}  

VLBI visibilities were analyzed via two standard pathways: the first analysis used only ``closure'' quantities, which provide immunity to station-based calibration errors, and the second analysis used ``self-calibration,'' which attempts to simultaneously solve for source structure and complex, time-dependent station gains.

\subsection{Fitting an Elliptical Gaussian Using Closure Amplitudes}
\label{sec::Closures}

For a closed triangle of interferometric baselines, the phase of the bispectrum (the directed product $V_{12} V_{23} V_{31}$ of the three complex visibilities $V_{ij}$ around the triangle) is immune to any station-based phase errors. This quantity is known as a ``closure phase.'' Likewise, closure amplitudes, such as $\left| (V_{12} V_{34})/(V_{13} V_{24}) \right|$, can be constructed for any quadrangle of sites and provide immunity to station-based gain amplitude errors \citep{TMS_2007}. We constructed closure amplitudes and phases from the phase-only calibrated data for each 10-second time segment. 

\begin{figure}[t]
\begin{center}
\includegraphics[width=1\columnwidth]{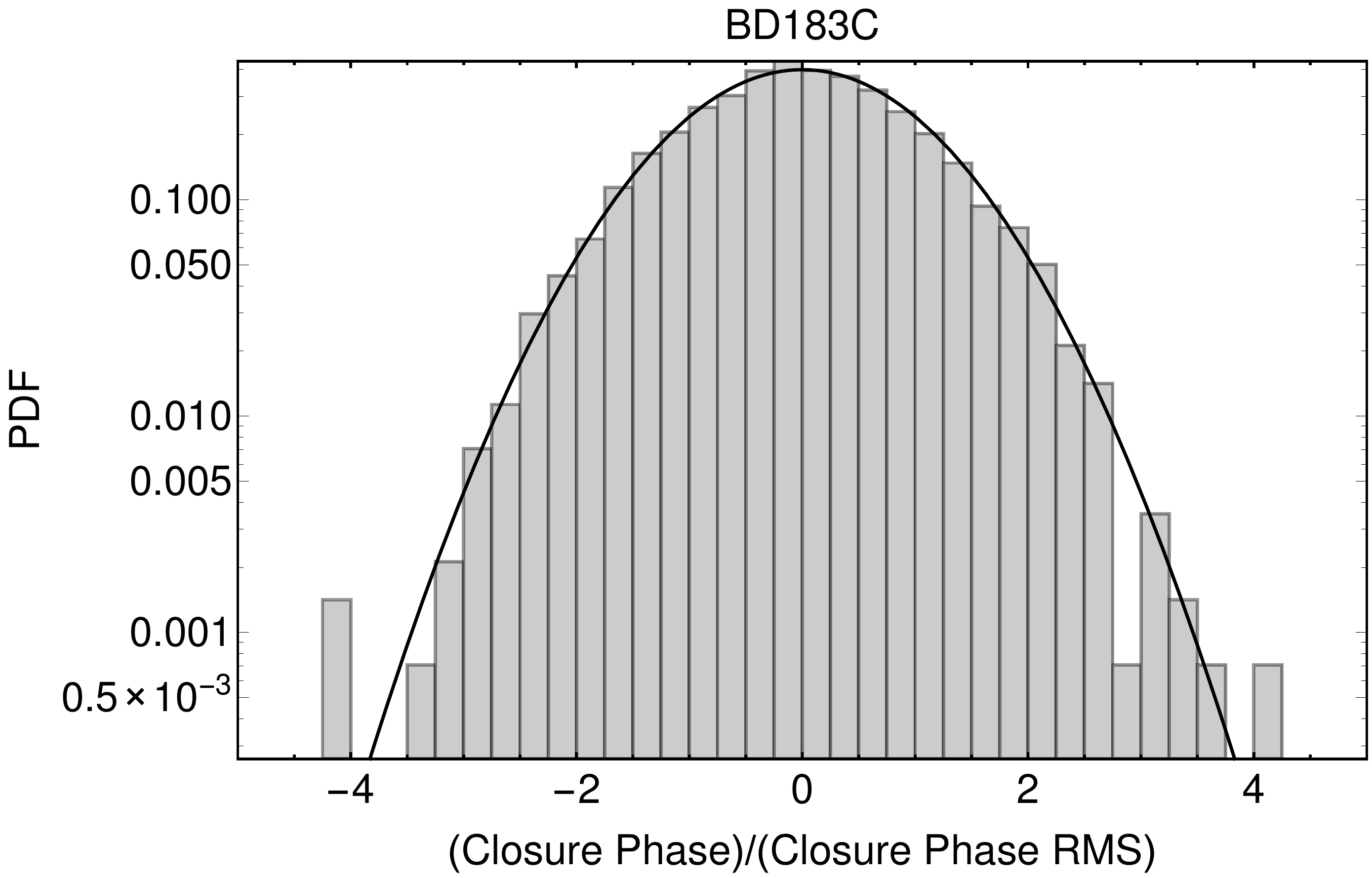}
\caption{\label{fig:clHist} Probability density function (PDF) of the
standardized closure phases on all triangles with baselines shorter
than $250~{\rm M}\lambda$. The solid line shows a fitted Gaussian, representing  
zero intrinsic closure phase and non-zero measurements entirely due to
thermal noise. We use this Gaussian fit to estimate the scaling factor relating each ``weight'' 
reported by AIPS for a complex visibility to the thermal noise.}
\end{center}
\end{figure}

Measured closure phases from both days are consistent with a zero-mean Gaussian distribution (see Figure \ref{fig:clHist}). We then fit the distribution of closure phases to calculate a single coefficient that converts AIPS weights $w_i$ to thermal noise $\sigma_i \propto 1/\sqrt{w_i}$ for each measurement. Because the atmospheric coherence time at $\lambda=3.5~{\rm mm}$ is only tens of seconds and coherent averages must be done over even shorter timescales to preserve the closure relationships discussed below, most of our measurements have only moderate signal-to-noise. For example, the median signal-to-noise in our two observing epochs was 8.3 and 7.2, respectively, for all detections, but ${\sim}10\%$ of detections have ${\rm SNR} < 3$. Both closure amplitudes and phases have markedly non-Gaussian errors in this regime, and closure amplitudes suffer a noise bias. For example, for a closure amplitude constructed from four visibilities that each have an SNR of 3, the average will be biased upward by $30\%$, and estimates of the closure amplitude uncertainty using high-SNR properties will be incorrect. For this reason, we derived the conversion between AIPS weights and thermal noise using closure phases with ${\rm SNR} > 3$, and we used Monte Carlo simulations to estimate the bias and uncertainties in our closure quantities.

\begin{figure}[t]
\begin{center}
\includegraphics[width=1\columnwidth]{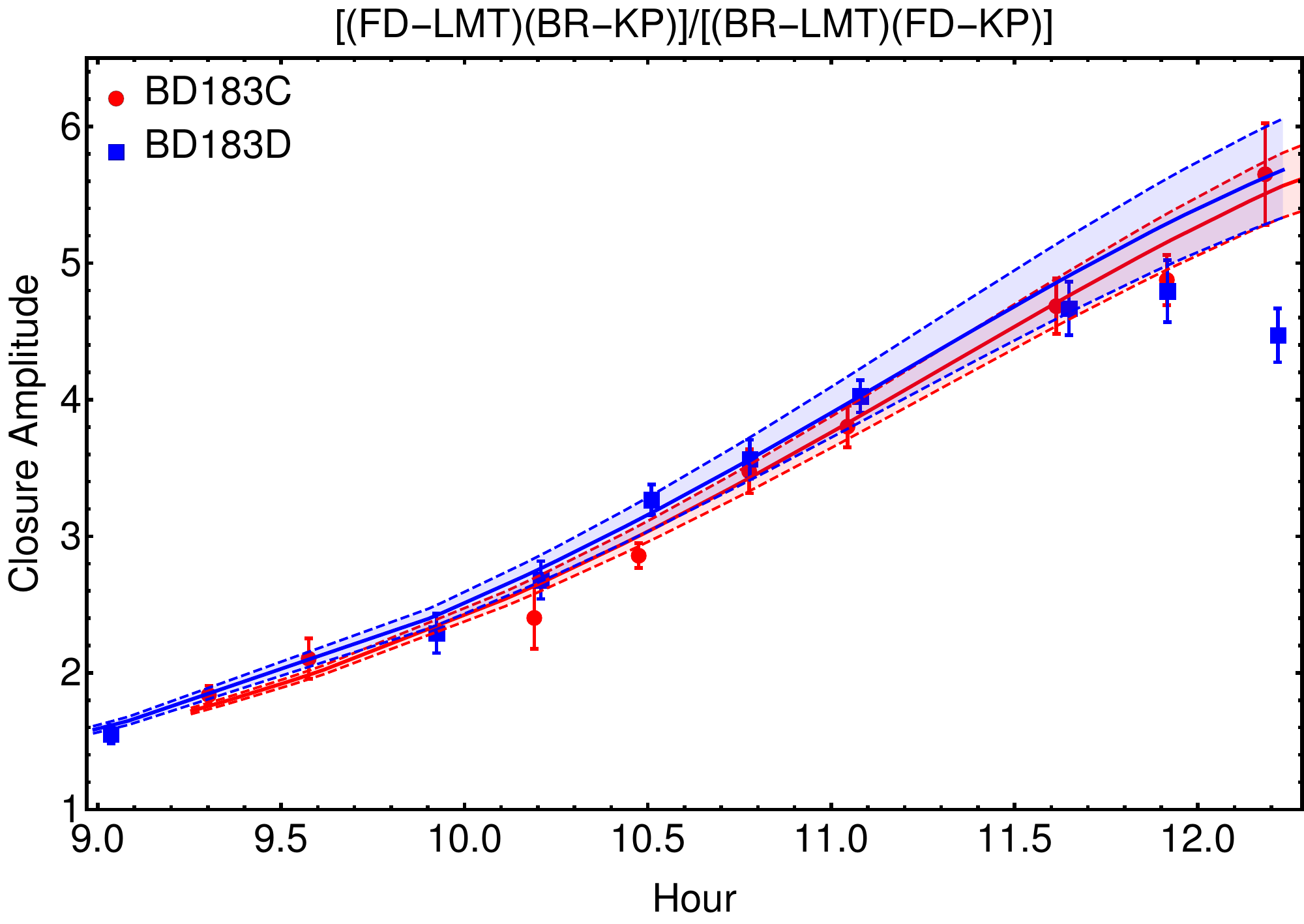}\\
\includegraphics[width=1\columnwidth]{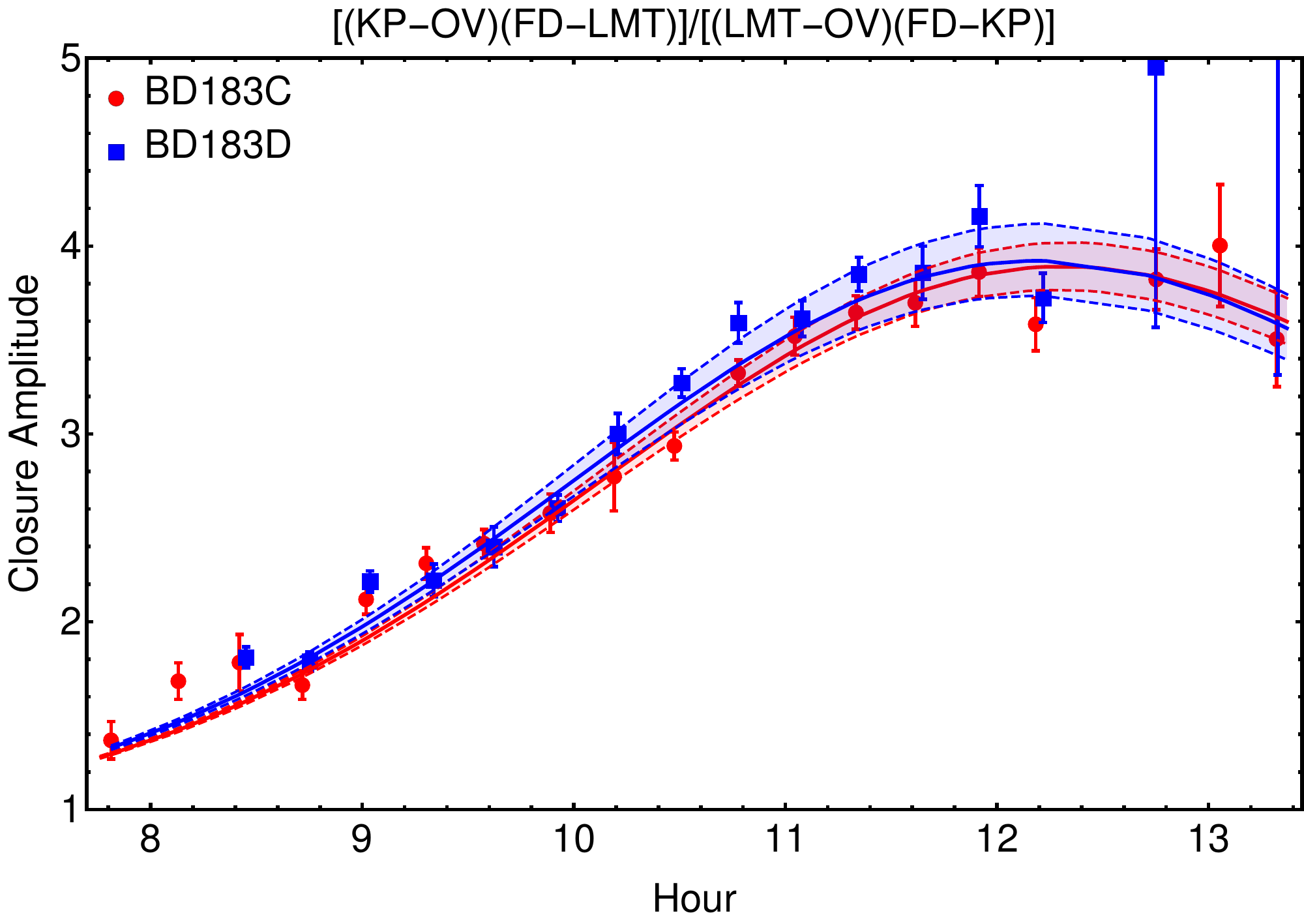}
\caption{\label{fig:clamp} Examples of closure amplitudes on two quadrangles. The points show scan-averaged closure amplitudes; the lines and shaded regions show the best-fit model from self-calibration in each epoch and $\pm 1\sigma$ uncertainty. 
}
\end{center}
\end{figure}

Even after averaging our closure phases over each epoch, they are still close to zero, consistent with an elliptical Gaussian structure. Consequently, for both epochs BD183C and BD183D, we performed a least-squares fit of elliptical Gaussian source models to the closure amplitudes (see Figure \ref{fig:clamp}). To avoid errors that were significantly non-Gaussian and the associated bias, we only used closure amplitudes constructed from visibilities that had ${\rm SNR} > 3$ in their 10-second coherent average for these fits.

The best-fit solutions have a reduced $\chi^2$ of $1.50$ for BD183C and $1.25$ for BD183D. These values are greater than unity at high significance, so to determine whether the excess can be entirely accounted for by the non-Gaussian closure amplitude errors, 
 we generated synthetic data sets for each epoch using the best-fit elliptical Gaussian model for the source.
We sampled the model on each baseline for which there was a detection, and added the expected amount of thermal noise to each sample. 
Finally, we calculated closure amplitudes for these synthetic data and used them to find the best-fit elliptical Gaussian.
 This procedure successfully reproduced the input model within the derived uncertainties and found a corresponding reduced $\chi^2$ of ${\sim}1.25$ in both epochs. Thus, the excess in our reduced $\chi^2$ is comparable to what is expected from the non-Gaussian errors on the closure amplitudes.

Unlike previous efforts \citep[e.g.,][]{Bower_2004,Shen_2005,Bower_2014}, we did not use the $\chi^2$ hypersurface to estimate parameter uncertainties in the fits to closure amplitudes. Several problems in this approach have been noted by \citet{Doeleman_2001}. Namely, because the closure amplitudes are not independent, a fixed increase $\Delta \chi^2$ does not accurately represent an expected confidence interval. As a trivial example of this, duplicating a data set will double the $\Delta \chi^2$ but obviously does not  constrain model parameters better. Because there are nominally ${\sim}N^4$ closure amplitudes for $N$ stations but only ${\sim}N^2$ visibilities and independent closure amplitudes, the redundant information can be substantial. Non-Gaussian noise, especially the high tail in the closure amplitude distribution, can also invalidate a standard $\chi^2$ approach. 

Instead, we estimated the uncertainty of the Gaussian parameters using a Monte Carlo simulation, independently fitting elliptical Gaussians to 20 different new data sets that each added additional thermal noise to the original complex visibilities with equal standard deviation to their original thermal noise before constructing the closure amplitudes for each set. We then report uncertainties given by the scatter in the fitted parameters. Note that because this procedure decreases the SNR of each measurement by a factor of $1/\sqrt{2}$, it conservatively estimates the parameter uncertainties. Table~\ref{tab::Gaussian_Fits_Scattered} gives our best-fit model and its associated uncertainty in each epoch.

\subsection{Fitting an Elliptical Gaussian Using Self-Calibration}
\label{sec::Self_calibration}

We also fit an elliptical Gaussian to the complex visibilities using ``self-calibration.'' This approach fits the Gaussian model simultaneously with time-dependent complex station gains. In this case, measurement uncertainties are described simply as additive complex Gaussian noise, and so there is neither bias nor non-Gaussian noise to contend with, even when the SNR is low. Thus, self-calibration can reliably utilize weaker detections than the closure-only analysis.

A concern for self-calibration is that the derived model can be heavily biased by the input self-calibration model (the initial guess for the source structure), especially if the minimization is not permitted to iteratively converge (\citet{Bower_2014} illustrates this unsurprising bias for self-calibration with a single iteration). A second concern is that parameters reported for self-calibration are often computed without accounting for the uncertainties in the self-calibration parameters -- the $\Delta \chi^2$ is explored over the space of model parameters while holding the best-fit self-calibration solution constant. Such estimates can significantly underestimate model parameter uncertainties.

We self-calibrated our data by independently deriving gain solutions for every 10-second integration. 
We confirmed that the self-calibration (iteratively) converged to the same solution regardless of the initially specified model. Specifically we checked convergence by comparing the results with two initial models: a point source and a 500~$\mu$as circular Gaussian source.

We then use the  $\chi^2$ hypersurface of both the self-calibration and elliptical Gaussian parameters to evaluate uncertainties in the model. We only included points with ${\rm SNR} > 3$ to avoid potentially spurious or corrupted detections. This restriction eliminates ${\approx}10\%$ of our data but only $<2\%$ of the LMT detections because of their higher SNR. The best-fit model in each epoch and the corresponding model uncertainties are given in Table~\ref{tab::Gaussian_Fits_Scattered}. 

We also repeated the estimate of uncertainties in the Gaussian model parameters while holding the self-calibration solution constant and equal to the best-fit self-calibration solution (this is the most straightforward self-calibration approach in AIPS, for instance). The derived Gaussian parameter uncertainties were a factor of ${\sim}6$ smaller for the major and minor axes, and were a factor of ${\sim}10$ smaller for the position angle, showing that the self-calibration uncertainties are a critical part of the error budget even when the self-calibration is allowed to iteratively converge.

{
\centering
\begin{deluxetable*}{l|cc|cc|ccc}
\tablecaption{Summary of Elliptical Gaussian Fits to 3-mm VLBI of \sgra.}
\tablehead{
\colhead{}  & \multicolumn{2}{c}{\bf BD183C} & \multicolumn{2}{c}{\bf BD183D} & \colhead{Doeleman+~($'$01)}  & \colhead{Shen+~($'$05)}  & \colhead{Lu+~($'$11)}\\ 
\colhead{} & \colhead{ Closure Amp. } & \colhead{ Self Calibration }  & \colhead{ Closure Amp. } & \colhead{ Self Calibration } & \colhead{(Self-Cal)} & \colhead{(Cl.~Amp.)} & \colhead{(Self-Cal)} 
}
\startdata
Maj.~Axis        & $214.9 \pm 4.0~\mu{\rm as}$ & $212.7 \pm 2.3~\mu{\rm as}$ & $217.7 \pm 5.0~\mu{\rm as}$ & $221.7 \pm 3.6~\mu{\rm as}$  & $180 \pm 20~\mu{\rm as}$   & $210^{+20}_{-10}~\mu{\rm as}$ & $210 \pm 10~\mu{\rm as}$ \\ 
Min.~Axis        & $139.0 \pm 8.1~\mu{\rm as}$ & $138.5 \pm 3.5~\mu{\rm as}$ & $147.3 \pm 8.0~\mu{\rm as}$ & $145.6 \pm 4.0~\mu{\rm as}$  & ---                                          & $130^{+50}_{-130}~\mu{\rm as}$   & $130 \pm 10~\mu{\rm as}$ \\
P.A.    & $80.8^\circ \pm 3.2^\circ$  & $81.1^\circ \pm 1.8^\circ$  & $80.2^\circ \pm 4.8^\circ$  & $75.2^\circ \pm 2.5^\circ$   & ---   & $79^{+12\hspace{0.2ex}\circ}_{-33}$   & $83.2^\circ \pm 1.5^\circ$ \\
Axial Ratio       & $1.55 \pm 0.08$             & $1.54 \pm 0.04$             & $1.48 \pm 0.07$             & $1.52 \pm 0.05$             & ---              & $1.62_{-0.6}^{+20}$ & $1.62 \pm 0.11$ 
\enddata
\label{tab::Gaussian_Fits_Scattered}
\tablecomments{Our elliptical Gaussian fits to the scattered image of \sgra\ at $\lambda=3.5~{\rm mm}$ and previously published values. 
Major and minor axes are given as the full width at half maximum (FWHM). \citet{Doeleman_2001} found that their data did not warrant an elliptical Gaussian model rather than a circular Gaussian; their quoted uncertainties include the effects from uncertainties in the self-calibration solution and from thermal noise. \citet{Shen_2005} only placed upper limits on the minor axis and did not measure anisotropy at high statistical significance. \citet{Lu_2011} reported fits and uncertainties from self-calibration and used the spread of fitted size among different epochs to estimate the overall uncertainty. However, their reported spread in fitted values from epoch to epoch did not include the two epochs for which an elliptical model is underdetermined.
Consequently, the uncertainties reported by \citet{Lu_2011} in minor axis size are likely too small by a factor of ${\sim}2-3$. 
Note that the axial ratio and its corresponding uncertainty was not reported in \citet{Shen_2005} or \citet{Lu_2011}; we derived these quantities using a skew normal distribution for the \citet{Shen_2005} results and a normal distribution for \citet{Lu_2011}, each with uncorrelated errors on the major and minor axes.\\
}
\end{deluxetable*} 
}

\subsection{Self-Calibration vs. Closure-Only Analysis}  

There has been considerable discussion in the literature about whether self-calibration or closure-only analysis is preferable for fitting Gaussian models to Sgr~A$^\ast$ \citep[e.g.,][]{Doeleman_2001,Bower_2004,Shen_2005,Bower_2014}. We have performed both analyses and found consistent results both in the best-fit models and for their associated parameter uncertainties when the self-calibration model uncertainties are properly taken into account. We do find that the self-calibration uncertainties are still smaller by a factor of ${\sim}2$, even after accounting for uncertainties in the self-calibration solution. Overall, our data suggest that 
both approaches should be used and checked for consistent results; differences may highlight problems in the assumptions for deriving the uncertainties of either model. 

\subsection{The Role of the LMT}

Prior attempts to constrain the minor (NS) axis size of \sgra\ have met with varied success. \citet{Shen_2005}, who analyzed closure amplitudes, could only determine an upper bound for the minor axis size; likewise, \cite{Lu_2011}, who self-calibrated to an elliptical Gaussian model, found that in 2 out of 10 observing epochs the elliptical model is underdetermined.  When LMT baselines are excluded from the analysis presented here, the results are similar. Specifically, even when including weak detections, self-calibration to the BD183C data without the LMT gave a minor axis size of $153 \pm 15~\mu{\rm as}$. However, in BD183D, the self-calibration finds a best-fit minor axis of $67_{-67}^{+40}~\mu{\rm as}$ (i.e., a size of zero is excluded at a significance of ${<}1\sigma$).  Likewise, in both epochs, fits using only closure amplitudes could only estimate an upper-bound for the minor axis size, so the self-calibration solutions must be interpreted with caution.  We then conclude that past measurements could only confidently measure an upper-bound for the minor axis size of the scattered image of \sgra\ at $\lambda=3.5~{\rm mm}$ in individual observing epochs. This analysis confirms that inclusion of the LMT baselines is essential to the robust determination of the intrinsic size in the N-S direction.  This result is unsurprising because the geographical location and size of the LMT significantly improves the north-south coverage and sensitivity of the VLBI array.

\section{Results}


The size of the scattering ellipse can be estimated based on the wavelength-dependent size of \sgra\ at wavelengths longer than a few cm. \cite{Bower_2006}  determined the normalization of the scattering law to be given according to $1.31\times 0.64$ mas cm$^{-2}$ at 78$^{\rm o}$ east of north. The uncertainties in these values are $\pm0.03$ mas in major axis size,  ${+0.04}$ and ${-0.05}$ mas in minor axis size and $\pm 1^{\rm o}$ in position angle. At 3.5 mm, this law gives  scattering sizes of $159.2\pm3.6~\mu{\rm as}$ and $77.8^{+4.9}_{-6.1}~\mu{\rm as}$ for major and minor axis, respectively.  This ellipse is  smaller that the ellipse measured at both epochs from closure quantities and self-calibration, which means that we are detecting the intrinsic structure of the source. However, \cite{Psaltis_2015} have also analyzed the set of measured sizes of \sgra\ and suggest that there are large systematic errors in the minor axis size.
Indeed, our comparison of self-calibration and closure-only results reinforces the suspicion that uncertainties derived in previous experiments may be systematically low. 

We deconvolve the measured ellipse with the scattering ellipse to determine the intrinsic size and orientation of \sgra. 
To properly account for the errors, we perform a Monte Carlo simulation. For this simulation we create 10 000 realizations of the observed ellipse, by taking independently a major axis size, a minor axis size and a position angle from Gaussian distributions with standard deviations equal to the errors given in Table \ref{tab::Gaussian_Fits_Scattered}. For each of these realizations, we similarly create a realization of the scattering ellipse, with parameters taken from Gaussian distributions that have a variance equal to the quadratic sum of the errors reported by \cite{Bower_2006} and the systematic errors by \cite{Psaltis_2015}. These systematic errors are 3\% in the major axis, 25\% in the minor axis, and 12\% in the position angle.  
We then take the deconvolution with the observed ellipse for each realization, and compute the ratio of major to minor axis, $A_{\rm int}$. 
The resulting distributions are symmetric Gaussians for the intrinsic major axes with means and standard deviations given in Table \ref{tab::IntrinsicSizes}. The distributions for minor axis, position angle, and axial ratio are non-Gaussian, so
we give for those the median and the 15.87th and 84.13th percentiles ($-\sigma$ and $+\sigma$) in Table \ref{tab::IntrinsicSizes}. We note that errors estimated using this approach are comparable to those derived by standard error propagation.  Within the accuracy of our measurements we do not see significant variations from one epoch to other in the intrinsic sizes of major and minor axis, and in position angle. For our two observations, we performed a weighted average of the closure  and self-calibrated intrinsic size estimates to arrive at an intrinsic ellipse of  $147\pm6~\mu{\rm as} \times 120^{+10}_{-13}~\mu{\rm as}$, at $88{^\circ}^{+7}_{-3}$ for the closure approach and $148\pm5~\mu{\rm as} \times 118^{+8}_{-10}~\mu{\rm as}$,  at $81{^\circ}\pm3{^\circ}$ for the self-calibration approach. The corresponding axial ratios of major to minor intrinsic size are $1.23^{+0.16}_{-0.09}$ and $1.26^{+0.14}_{-0.08}$, respectively. Considering that the Schwarzschild radius ($R_{\rm sch}$) for a black hole of mass $4.3\times10^{6}~\rm{M}_\odot$ \citep{Gillessen_2009} at a distance of 8.34 kpc \citep{Reid_2014} is $10.2~\mu{\rm as}$, the intrinsic angular sizes can be translated into physical sizes. The resulting values are $14.4\pm0.6~R_{\rm sch} \times 11.8^{+1.0}_{-1.3}~R_{\rm sch}$ for the closure approach and $14.5\pm0.5~R_{\rm sch}\times11.6^{+0.8}_{-1.0}~R_{\rm sch}$ for the self-calibration approach. 

{
\centering
\begin{deluxetable*}{l|cc|cc|ccc}
\tablecaption{Summary of Intrinsic Sizes of Sgr~A$^\ast$ at 3.5~mm.}
\tablehead{
\colhead{}  & \multicolumn{2}{c}{\bf BD183C} & \multicolumn{2}{c}{\bf BD183D} & \colhead{Doeleman+~($'$01)}  & \colhead{Shen+~($'$05)}  & \colhead{Lu+~($'$11)}\\ 
\colhead{} & \colhead{ Closure Amp.  } & \colhead{ Self Calibration }  & \colhead{ Closure Amp.  } & \colhead{ Self Calibration } & \colhead{(Self-Cal)} & \colhead{(Closure Amp.)} & \colhead{(Self-Cal)} 
}
\startdata
Maj.~Axis & $145 \pm 9~\mu{\rm as}$  & $142 \pm 7~\mu{\rm as}$  & $149\pm9~\mu{\rm as}$  & $155 \pm 8~\mu{\rm as}$    & $82\pm{46}~\mu{\rm as}$  & $136^{+32}_{-18} ~\mu{\rm as} $   & $ 139\pm17 ~\mu{\rm as}$ \\
Min.~Axis & $114^{+14}_{-19}~\mu{\rm as}$ & $113^{+11}_{-17}~\mu{\rm as}$ & $124^{+13}_{-17}~\mu{\rm as}$ & $122^{+11}_{-16}~\mu{\rm as}$  &    ---                                   & $104^{+65}_{-164} ~\mu{\rm as} $  & $ 102\pm21 ~\mu{\rm as}$  \\
P. A.         & $88{^\circ}^{+9}_{-4}$         & $89{^\circ}^{+10}_{-4}$          &  $87{^\circ}^{+13}_{-4}$        &  $69{^\circ}^{+3}_{-5}$          &    ---                                  &  $82^{+15^\circ}_{-34^\circ}$          & $95{^\circ}\pm10{^\circ}$  \\
Axial ratio & $1.27^{+0.26}_{-0.15}$                & $1.25^{+0.22}_{-0.12}$                      & $1.20^{+0.21}_{-0.12}$                 &  $1.27^{+0.19}_{-0.12}$                     &    ---                                 & 1.31$^{+0.87}_{-2.07}$                   & $1.36\pm0.33$\\
\enddata
\label{tab::IntrinsicSizes}
\tablecomments{We apply the same deconvolution scheme to the measured sizes by \citet{Doeleman_2008}, \citet{Shen_2005}, and \citet{Lu_2011} to arrive at the values listed in this table.
Notice that the measurement by  \citet{Lu_2011} resulted from an average over 8 epochs, while here we are able to determine the  intrinsic size and orientation at individual epochs. 
}
\end{deluxetable*} 
}

We now use past measurements of the scattered image at  1.3, 7, and 13.5 mm to study the dependence of the intrinsic size as a function of wavelength. We again use the kernel from \citet{Bower_2006} to remove the effects of  scattering and determine the intrinsic size of major and minor axis at these wavelengths. 

At 7 mm, the scattered two-dimensional image has been reported by \citet{Bower_2014}, \citet{Lu_2011}, and \citet{Shen_2005}. 
At 13.5 mm there are measurements by  \citet{Bower_2004}, and \citet{Lu_2011}. We follow the approach described above for deconvolution of these five size measurements with the scattering ellipse. At 1.3 mm, the (NS) apparent size is not well constrained \citep{Doeleman_2008}, so the scattered source  at this wavelength is assumed to be given by a circular Gaussian distribution.  We find that  intrinsic sizes  at a given wavelength from measurements by different authors are consistent within the errors. 

We note that, when the uncertainties reported by \cite{Psaltis_2015} are included in the error budget of the scattering kernel,  the axial ratio of intrinsic sizes at  7 mm is not statistically significant. Specifically, an axial ratio of $2.78^{+4.79}_{-4.94}$ is found and then this measurement should be taken cautiously.

\begin{figure}[h!]
\begin{center}
\includegraphics[width=\columnwidth]{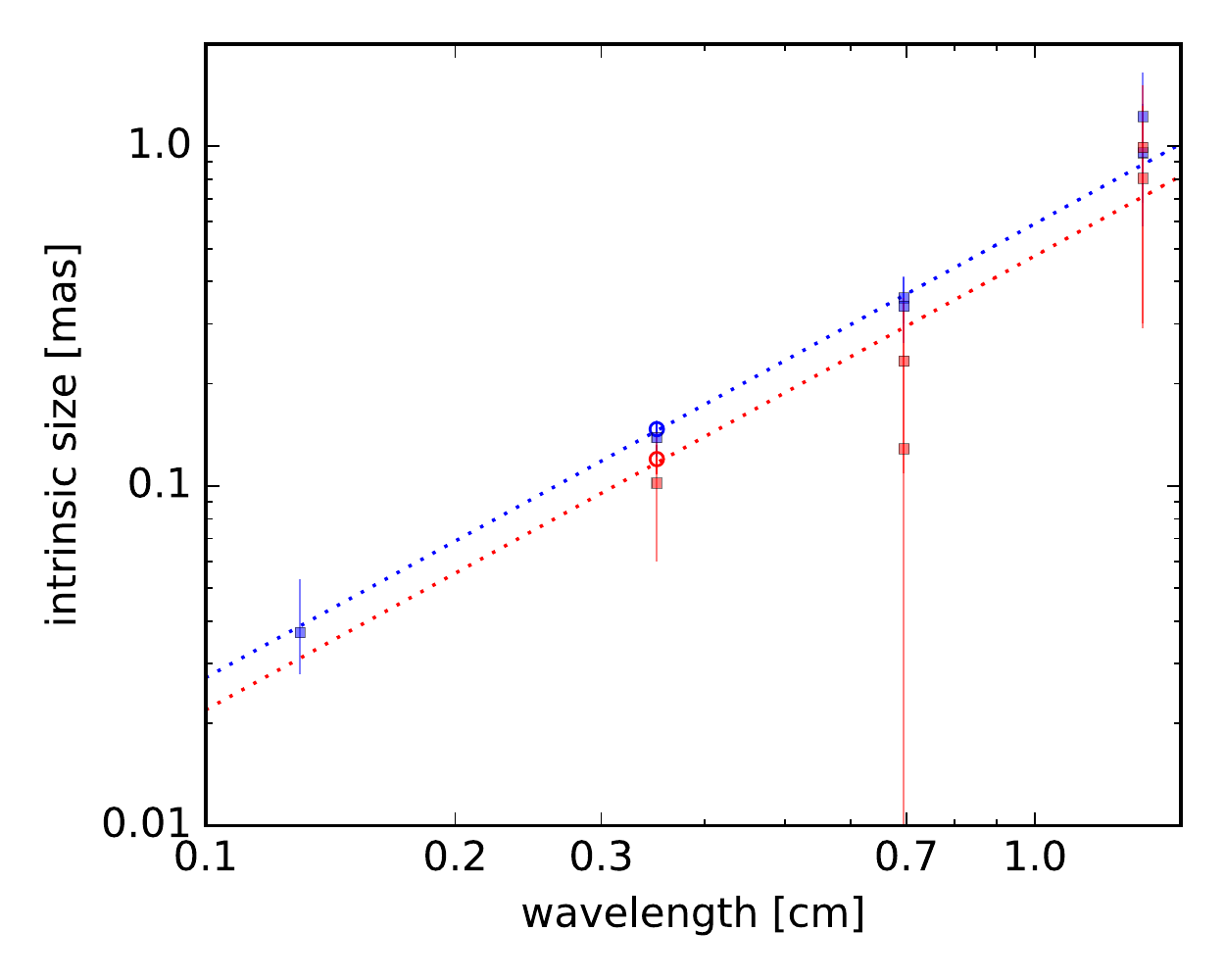}
\caption{\label{fig:size_vs_wl} Plot of intrinsic major (blue) and minor (red) axis size versus wavelength. The open circles at 3.5 mm correspond to the measurements reported in this work from closure approach. The squares  at 1.3~mm, 3.5~mm, 7~mm and 13.5~mm were obtained by reanalyzing the measurements from  \citet{Doeleman_2008}, \citet{Bower_2014}, \citet{Bower_2004}, and \citet{Lu_2011}. The dotted lines represent a fit to a power-law trend with common index of $1.34\pm0.13$ for both major and minor axes.
}
\end{center}
\end{figure}

To  investigate if the axial ratio scales with wavelength, we show in Figure \ref{fig:size_vs_wl}
the intrinsic sizes derived from the measurements  by \citet{Doeleman_2008} at 1.3~mm, \citet{Bower_2014} at 7 mm, \citet{Bower_2004} at 13.5 mm, as well as  the measurements from \citet{Lu_2011} at  3.5, 7 and 13.5 mm, where we have multiplied the minor axis uncertainty at 3.5~mm by a factor of two. Our weighted averages of sizes  derived from the closure approach at 3.5 mm using the new observations presented here are also shown as open circles. 

Assuming that the data can be represented by a $\lambda^{\beta}$ law, we performed a weighted least-squares linear fit to all measurements obtaining $\beta = 1.34\pm0.13$. If the power-law indices for major and minor axes are allowed to differ, the respective fits give $\beta = 1.35\pm0.14$ and $\beta = 1.26\pm0.38$. The errors in the power law indexes are taken from the diagonal entries of the  covariance matrix constructed for the fits. Hence, within the errors of the measurements, the intrinsic size of the major and minor axes follow the same power law. More precise measurements at wavelengths other than 3.5 mm are necessary to enable a robust fit from the minor-axis data alone and an investigation of the  dependence of the intrinsic shape on wavelength.

The observed size at  3.5~mm also gives an absolute upper limit on the scatter broadening along the minor axis. Our measurements at both epochs are only $1.4-1.7\sigma$ above the minor axis suggested  by \citet{Psaltis_2015} at 3.5~mm, significantly constraining the scattering kernel.

\section{Discussion}

\subsection{Effects from Refractive Scattering}  

The ``blurring'' from interstellar scattering that causes the $\lambda^2$ scaling of the scattered image of \sgra\ at wavelengths longer than a few centimeters is an ensemble-average effect and so only strictly applies when the scattered image is averaged over a long period of time. Diffractive scattering of the intrinsic image with an elliptical Gaussian
kernel does not affect closure phase \citep{Fish_2014}. However, within individual observing epochs, refractive scattering causes the image to become fragmented and it \emph{does} introduce stochastic non-zero closure phase variations \citep{Johnson_Gwinn_2015}. The imprint of these stochastic fluctuations can then be used to constrain properties of both the intrinsic source and the turbulence in the scattering material \citep{Gwinn_2014}.

\begin{figure}[t]
\begin{center}
\includegraphics[width=1.0\columnwidth]{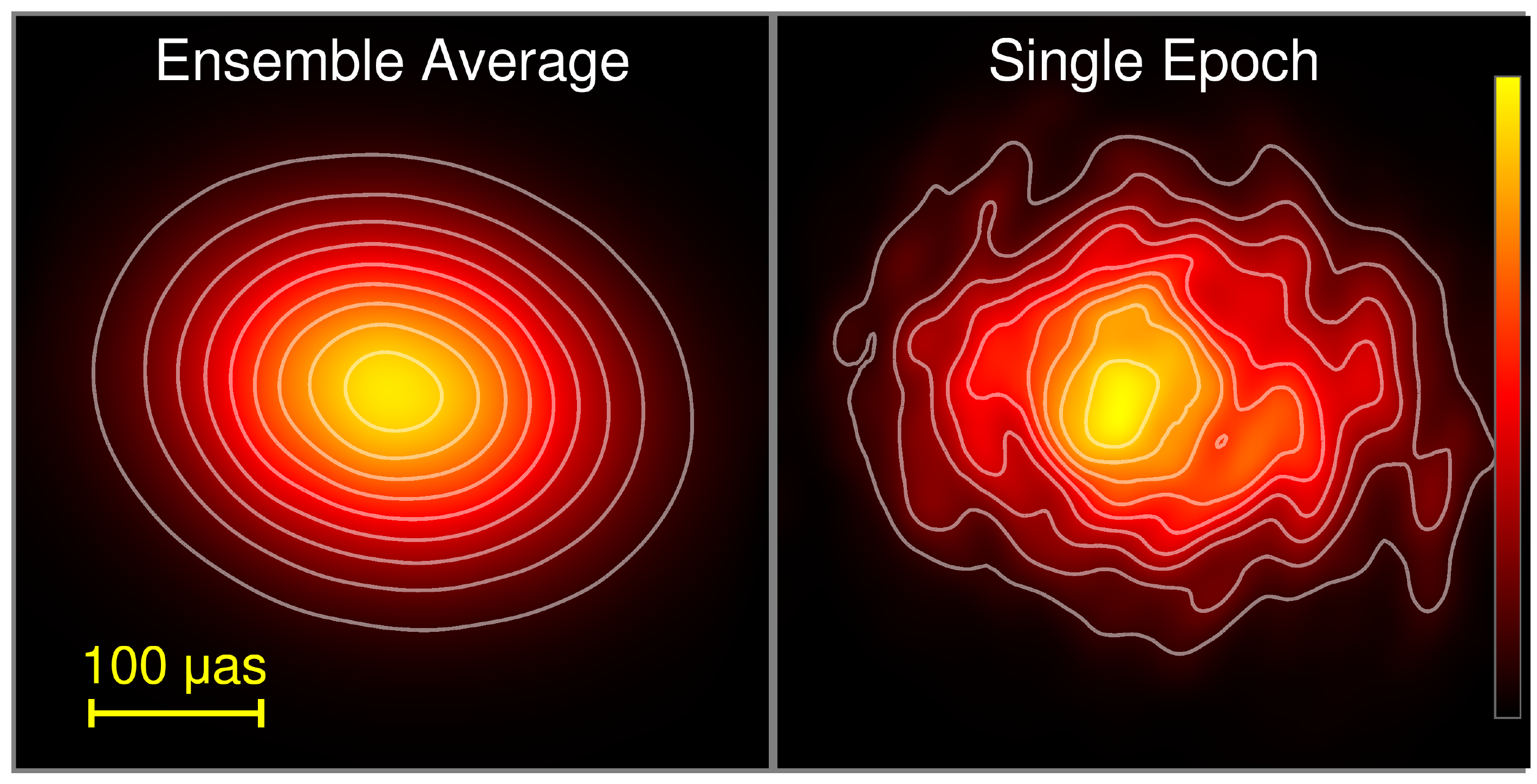}
\caption{Simulated scattered images of Sgr A$^\ast$ at $\lambda=3.5~{\rm mm}$; color denotes brightness on a linear scale, shown at the far right, and image contours are 10\% to 90\% of the peak brightness, in steps of 10\%.  The intrinsic source is modeled as a circular Gaussian with a FWHM of $130~\mu{\rm as}$; the ensemble-average scattered image has a FWHM of $(206~\mu{\rm as}) \times (151~\mu{\rm as})$. The left image shows an approximation of the ensemble-average image, obtained by averaging 500 different scattering realizations. This image illustrates the ``blurring'' effects of scattering when averaged over time. The right image shows the appearance for a single epoch, which exhibits scattering-induced asymmetries that would persist over a characteristic timescale of approximately one week. Each image has been convolved with a $20~\mu{\rm as}$ restoring beam to emphasize the features that are potentially detectable at $\lambda=3.5~{\rm mm}$.
}
\label{fig::Scattered_Image}
\end{center}
\end{figure}

\begin{figure*}[t]
\begin{center}
\includegraphics[width=0.32\textwidth]{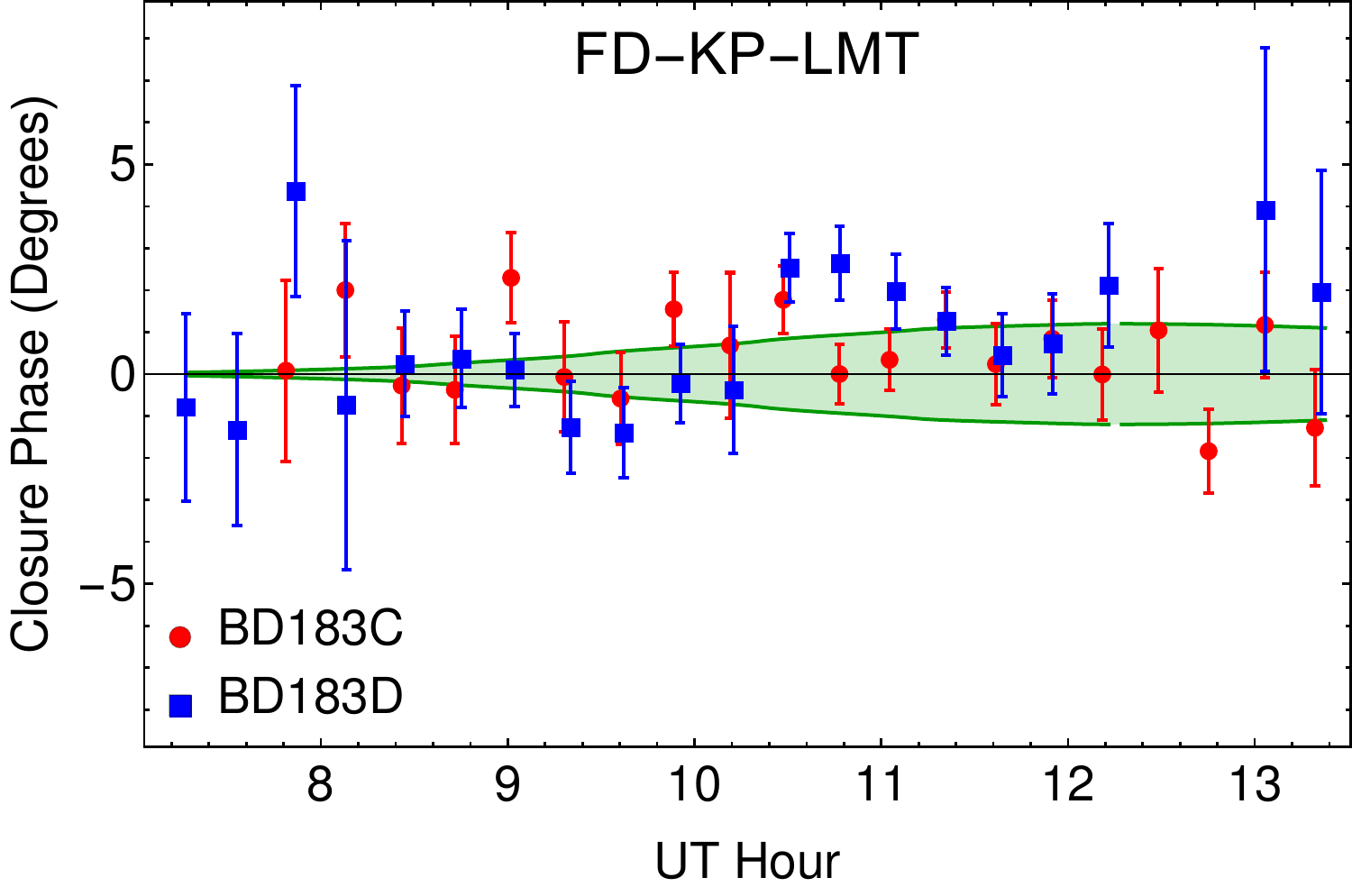}\hfill
\includegraphics[width=0.32\textwidth]{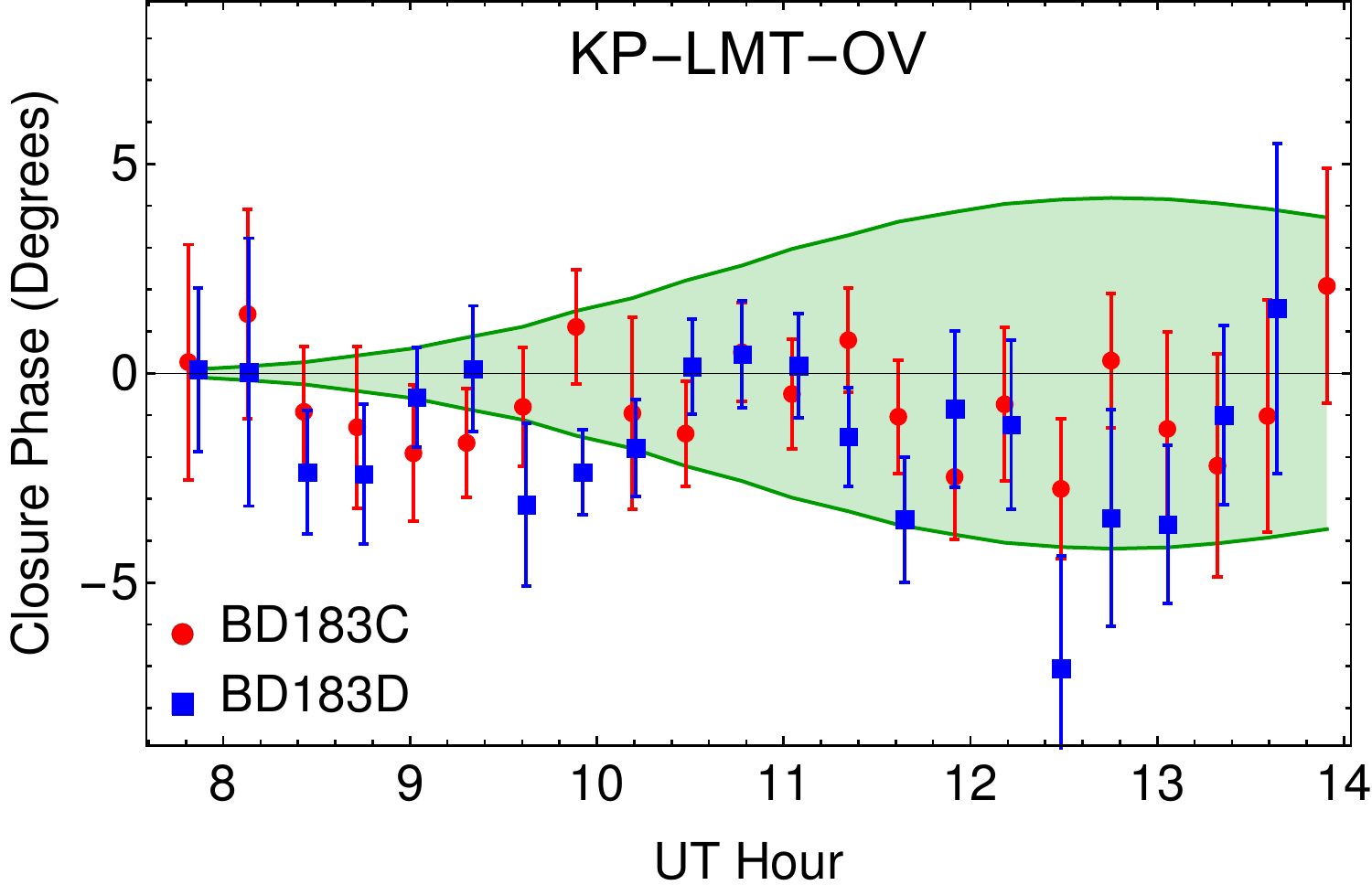}\hfill 
\includegraphics[width=0.32\textwidth]{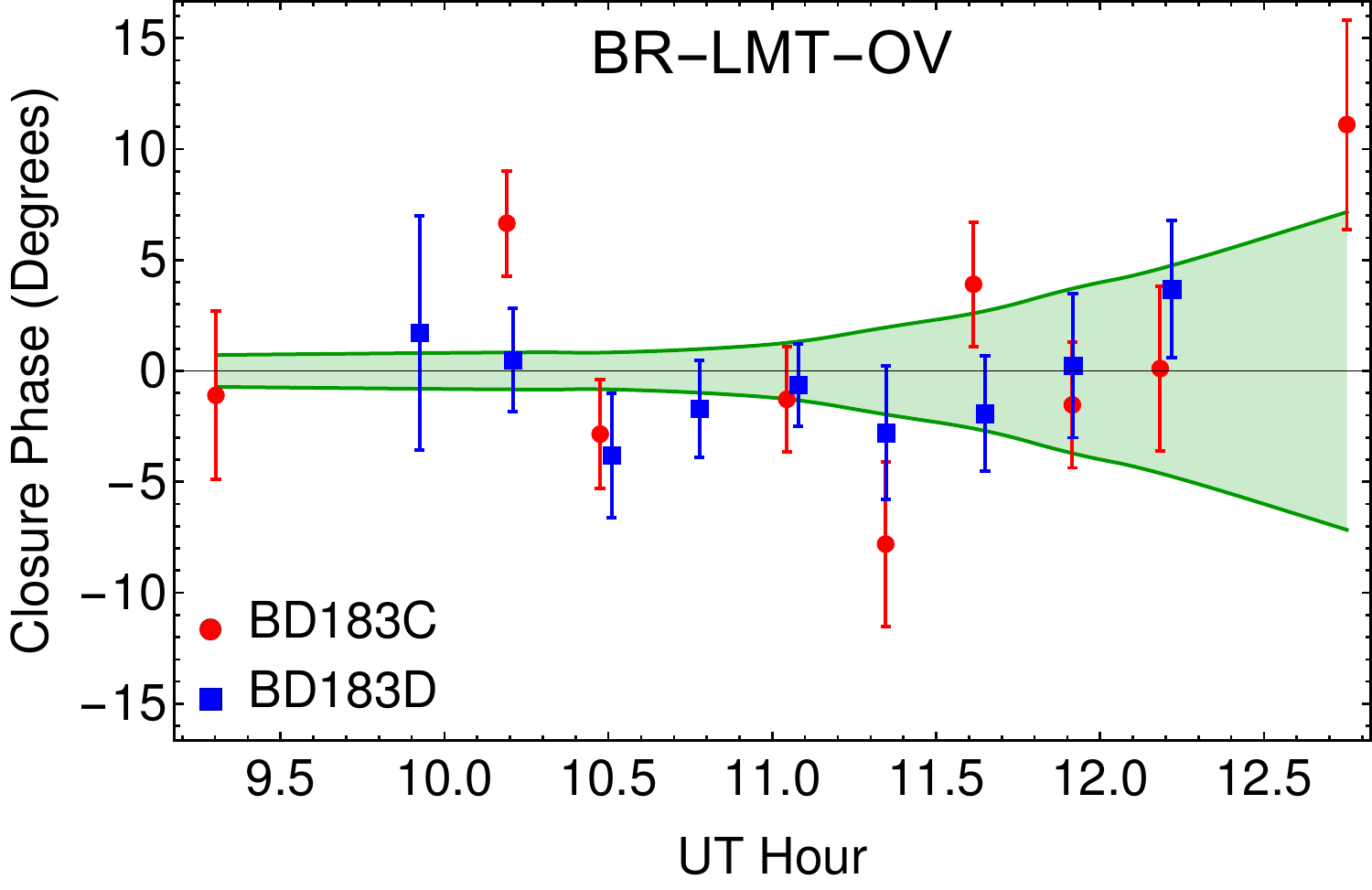}
\caption{Measured closure phases in each observing epoch as a function of time on three representative baseline triangles. The ${\pm} 1\sigma$ range of closure phase expected from refractive scattering of a $130~\mu{\rm as}$ circular Gaussian source is shown as the green shaded region of each plot. 
When the two epochs are combined, the average closure phases of the FD-KP-LMT ($0.67^\circ \pm 0.17^\circ$) and KP-LMT-OV ($-1.02^\circ \pm 0.24^\circ$) triangles are each non-zero at a significance of ${\sim}4\sigma$. However, these values are consistent with the expected closure phase excursions introduced by refractive scattering. Note that the scattering-introduced closure phases are largest when the visibility amplitudes are smallest, so the largest non-zero closure phases are also the most difficult to detect. 
}
\label{fig::CPhase_wRefractive}
\end{center}
\end{figure*}  

Refractive scattering causes flux modulation and  positional variation (image wander) at scales smaller than scattered size \citep{RCB_1984,Blandford_Narayan_1985,CPL_1986,Narayan_1992}. On baselines that are long enough to resolve the ensemble-average image, the refractive scattering introduces small-scale power from substructure that affects interferometric visibilities and which can be estimated analytically \citep{Narayan_Goodman_1989,Goodman_Narayan_1989,Johnson_Gwinn_2015}. However, effects from refractive scattering on closure amplitudes and closure phases for baselines that weakly or moderately resolve the image are difficult to estimate analytically. For this reason, we use numerical simulations of the refractive scattering to estimate the expected effects on our measurements. 

Following the methodology outlined in \cite{Johnson_Gwinn_2015}, we generated an ensemble of 500 scattered images of a circular Gaussian source with an intrinsic FWHM of $130~\mu{\rm as}$. For each image, we generated a scattering screen with $2^{13} \times 2^{13}$ correlated random phases corresponding to a Kolmogorov spectrum for the turbulence, and we determined the strength of the scattering by extrapolating the frequency-dependent angular size from longer wavelength measurements \citep{Bower_2006}. We assumed a scattering screen, placed at a distance of 5.8~kpc from the Galactic Center, as inferred by the combination of angular and temporal broadening from the Galactic Center magnetar \citep{Spitler_2014,Bower_2014_Magnetar}.  Refractive effects are, however, rather insensitive to the placement of the screen, with their strength scaling with $D^{-1/6}$, where $D$ is the observer-screen distance \citep{Johnson_Gwinn_2015}. Figure~\ref{fig::Scattered_Image} shows an example image from these scattering simulations. 

Each screen phase ``pixel'' had a linear dimension of approximately $0.5~\mu{\rm as}$, corresponding to $2 \times 10^5~{\rm km}$ which is still insufficient to resolve the phase coherence length, $r_0$, of the scattering screen, which is $(1200~{\rm km}) \times (2400~{\rm km})$ as determined by the angular size, $\theta_{\rm scatt}$ of the scattering kernel ($r_0 \sim \lambda/\theta_{\rm scatt}$). Because of this limitation, we set the inner scale, $r_{\rm in}$, of the scattering to be equal to the pixel resolution to ensure that the unresolved phase variations were smooth. For this reason, our simulations have slightly more refractive noise than expected, by a factor of ${\approx} (r_{\rm in}/r_0)^{1/6} \sim  1.6$, where $r_0$ is the phase coherence length of the scattering along the major axis. We divide the fluctuations of our simulations by this correction factor to derive comparisons with data. 

Our simulations gave a root-mean-square flux modulation of 6.6\%, which is reasonably close to the result from analytic calculations of $5.6\%$.
They also predict fractional modulation of the major and minor axes of the measured image of $3.1\%$ and $1.5\%$, respectively, or about $7~\mu{\rm as}$ for the major axis and $2~\mu{\rm as}$ for the minor axis. The expected fluctuation in the position angle of the scattered image is $2.0^\circ$. These fluctuations are potentially detectable among a set of multiple epochs when the LMT participates in VLBI with the VLBA. However, because our two observing epochs with the LMT are on consecutive days and the scattering likely evolves on a timescale of a week, the inter-epoch consistency in our measured parameters (see Table~\ref{tab::Gaussian_Fits_Scattered}) is expected. The timescale for the stochastic fluctuations to evolve is approximately given by the transverse size of the scatter-broadened image at the location of the scattering material divided by the transverse velocity of the scattering material \citep{Johnson_Gwinn_2015}. 
Assuming a transverse velocity of 50~km/s, we derive a characteristic timescale of approximately two weeks for the refractive scattering to evolve.

The fluctuations in visibility phase on each baseline are primarily determined by the visibility amplitude on that baseline. For an ensemble-average normalized visibility amplitude of $0.1 \lsim |V| \lsim 0.5$, the phase fluctuations in our numerical simulations are approximately $0.05/|V|~({\rm radians})$ for long east-west baselines and $0.03/|V|~({\rm radians})$ for long north-south baselines. However, because the phase fluctuations are correlated on similar baselines, the closure phase fluctuations are not well-approximated by the quadrature sum of these fluctuations. For example, phase fluctuations from image wander are entirely canceled in closure phase. 

Figure~\ref{fig::CPhase_wRefractive} compares our measured closure phases as a function of time on three representative baseline triangles with the root-mean-square fluctuations expected from refractive scattering. Our data exhibit some non-zero closure phases at high statistical significance (${\gsim}4\sigma$), but these values are consistent with being introduced by the scattering. Thus, while we find evidence for non-zero closure phases, we do not find evidence for \emph{intrinsic} non-zero closure phases. With additional observing epochs, the level of closure phase fluctuations could be used to constrain the scattering kernel and intrinsic structure of \sgra\ without relying on extrapolating the scattering kernel from longer wavelengths.


\subsection{Constraints on the Stratified Emission Structure of \sgra}

Our measurement of intrinsic source size at 3.5 mm and the $\lambda^\beta$ intrinsic size scaling provides a crucial constraint for any model of the emission from \sgra. Models that successfully reproduce the radio properties of \sgra\ usually separate outflow from accretion inflow for the emission. On the one hand, Radiatively Inefficient Accretion Flow (RIAF) models \citep[e.g.,][]{Yuan_2003,Broderick_2009} suggest  that the submillimeter emission stems from thermal electrons in the inner parts of the accretion flow. The intrinsic intensity profile, however, cannot be well described by a Gaussian distribution \citep{Yuan_2006}. 
In the semi analytical jet model of \citet{Falcke_2000}, on the other hand, the intrinsic structure is comprised of two components, the jet and the nozzle, whose length and width at 3.5 mm are $\sim160~\mu$as (15.7~R$_{sch}$) and $\sim48~\mu$as (4.7~R$_{sch}$), respectively. At this frequency, the nozzle dominates the millimeter emission. 
In this model, the jet length scales as $\lambda^{m}$, with $m\sim 1$, and the axial ratio of major to minor axis of the jet is $\sim 3$ at 3.5 mm. While our results support a power-law dependence of the intrinsic size close to 1, we have found a somewhat symmetric  
deconvolved size, that does not agree with the intrinsic anisotropic structure predicted by such jet model. 

More sophisticated models, in which jets are coupled to a RIAF, are equally successful in explaining the spectrum of Sgr A$^\ast$.  \citet{Moscibrodzka_2014} conclude that the radio appearance is  dominated by the outflowing plasma; however, the geometry of the emitting region depends on model parameters, such as electron temperature in the jet and accretion disk, inclination angle of the jet, and the position angle of the black hole spin axis. Nevertheless, their best (bright jet) models   are within the size constraint imposed by our measurements at 3.5 mm. 

To unambiguously distinguish between the various models more accurate closure phase measurements are needed.  In addition, multi-epoch observations will be essential to unambiguously distinguish between intrinsic structure and refractive substructure from interstellar scattering.  The LMT has recently joined the EHT for 1.3~mm VLBI observations of \sgra\ and ALMA is planned to do so in the near term. Since at 1.3~mm the source structure is less contaminated by scattering, the EHT, when completed,  will  enable image reconstruction of the source.

\section{Summary}

We have used VLBI to study \sgra\ at 3.5-mm wavelength. Our results are the first to use the LMT as part of a VLBI network, providing significant improvements to the VLBA, especially in the north-south array coverage. We find that the image of \sgra\ at this wavelength is well characterized as an elliptical Gaussian, and we determine a robust measurement of the intrinsic size at this wavelength separately in two observing epochs. When our data are analyzed without including the LMT, we are unable to meaningfully constrain the intrinsic north-south structure  because the LMT adds the critical north-south baseline coverage. We also find that previous experiments reported significantly underestimated uncertainties in the minor axis size, principally because they did not considerer  the systematic errors in the scattering kernel. 
Our data show non-zero closure phases in \sgra, but we demonstrate that these values are consistent with being introduced by refractive scattering in the ionized interstellar medium; they do not yet provide evidence for asymmetric intrinsic structure at 3.5~mm wavelength.  Our measurements provide guidance for simulations and theories that describe the energetic accretion and outflow from \sgra, and they highlight the importance of refractive interstellar scattering for understanding the intrinsic structure of \sgra\ with short-wavelength VLBI imaging.

\acknowledgements{
G.\ N.\ O.-L., L.\ L., J.\ L.-T and A.\ H. acknowledge the financial support  of CONACyT, Mexico. G.\ N.\ O.-L., L.\ L.,  and A.\ H. also acknowledge DGAPA, UNAM, for financial support.  M.\ J. and S.\ D. acknowledge the Gordon and Betty Moore Foundation for financial support of this work through grant GBMF-3561. This work was supported by US National Science Foundation grants AST-1310896, AST-1211539 \& AST-1337663. V.L.F.  acknowledges support from the National Science Foundation. G.\ N.\ O.-L. is grateful to L.\ F.\ Rodr{\'{i}}guez for sharing his deconvolution script and for assistance with its use. 
{\it Facilities:} \facility{VLBA}, \facility{LMT}

\ \\

\bibliography{LMT_3mm_SGRA_revised.bib}

\end{document}